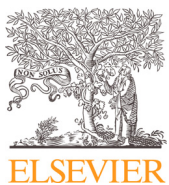
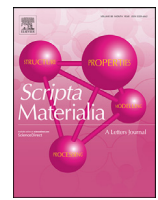

# Revealing atomic-scale vacancy-solute interaction in nickel

Felipe F. Morgado[a,*], Shyam Katnagallu[a,b], Christoph Freysoldt[c], Benjamin Klaes[c], François Vurpillot[a,c], Jörg Neugebauer[c], Dierk Raabe[a], Steffen Neumeier[d], Baptiste Gault[a,e], Leigh T. Stephenson[a]

[a] *Max-Planck-Institut für Eisenforschung, Max-Planck-Str. 1, 40237, Düsseldorf, Germany*
[b] *now at Karlsruhe Institut für Technologie (north campus), Hermann-von-Helmholtz-platz 1, 76344, Eggenstein-Leopoldshafen, Germany*
[c] *Groupe Physique des Matériaux, Université de Rouen, Saint Etienne du Rouvray, Normandie 76800, France*
[d] *Materials Science and Engineering, Institute 1, Friedrich-Alexander-Universität Erlangen-Nürnberg, Erlangen, Germany*
[e] *Department of Materials, Royal School of Mines, Imperial College London, London, SW7 2AZ, United Kingdom*

A B S T R A C T

It is widely accepted that the different types of crystalline imperfections, such as vacancies or dislocations, greatly influence a material's physical and mechanical properties. However, imaging individual vacancies in solids and revealing their atomic neighborhood remains one of the frontiers of microscopy and microanalysis. Here, we study a creep-deformed binary Ni-2 at.% Ta alloy. Atom probe tomography reveals a random distribution of Ta. Field ion microscopy, with contrast interpretation supported by density-functional theory and time-of-flight mass spectrometry, evidences a positive correlation of Ta with vacancies, supporting positive solute-vacancy interactions previously predicted by atomistic simulations.

© 2021 Acta Materialia Inc. Published by Elsevier Ltd. All rights reserved.

The direct characterization of individual vacancies is one of the final frontiers of microscopy and microanalysis. Vacancies are involved in numerous materials processes, from diffusion to deformation, and the inability to precisely characterize a population of vacancies and their immediate neighborhood results in mechanisms being inferred from indirect evidence, leaving much room for interpretation and hence improvements. Field ion microscopy (FIM), the predecessor to atom probe tomography (APT), is likely the only technique allowing for imaging materials in three-dimensions with true atomic-resolution, including vacancies. FIM was developed in the 1950s and was the first technique to image individual atoms at a metal surface [1]. Through field ionization of an imaging gas near the sharp needle specimen's surface, FIM provides an atomically-resolved projection of the surface with near 100% imaging efficiency. The contrast in FIM images is associated with different chemical elements and local curvatures [2], making it impossible to unequivocally assign an elemental nature to each of the imaged atoms. Yet FIM was used to provide critical insights into crystallographic defects such as vacancies [3,4,5,6], dislocations [7,8,9], or grain boundaries [9,10], but studying the atomic-scale interactions between defects and solute elements within a material remains extremely challenging. In order to overcome this limitation, atom probe tomography (APT) has been extensively used [11,12], but at the cost of a lower spatial resolution. Recently, analytical FIM (aFIM), which combined FIM and time-of-flight mass spectrometry, demonstrated superior capabilities to APT and revealed *Re*-segregation to edge dislocations in the analysis of a Ni-*Re* binary alloy, used as a solid solution model substance to Ni-based superalloys [13].

Ni-based superalloys are critical engineering materials for load-bearing applications at high-temperature (e.g., 1100 °C) in harsh gaseous environments, enabling turbines for air traffic or large-scale energy conversion [14,15,16]. More specifically, single crystals are used in the most demanding environments to avoid failure by damage accumulation at grain boundaries [17,18]. Numerous studies have focused on understanding the deformation mechanisms, particularly for creep [19,20,21]. At high temperatures, creep occurs mainly through dislocation slip and dislocation climb [22,23]. The latter consists in the emission or absorption of vacancies by the core of an edge dislocation that will climb in the normal direction to its glide plane. Rhenium is the best solid solution strengthening element in Ni-based superalloys at high temperatures due to segregation to dislocations, imposing a dragging force [24]. However, *Re* is an expensive and rare element and, depending on the alloy, can represent half the cost of the raw material to produce a turbine blade [25]. Meanwhile, at lower temperatures other elements give a better solid solution strengthening effect. Tantalum, for instance, provides higher strength in the temperature range 800–1000 °C at an applied strain rate of $10^{-4}$ s$^{-1}$ [26]. At these temperatures and strain rates, the diffusion does not play a significant role, and the paraelastic interaction is the dominant solid solute hardening mechanism [26]. At lower strain/creep rates, the temperature where Ta is the better solid strengthener than $Re$ is shifted to lower temperatures. Although, Ta partitions predominantly to the Ni3Al $\gamma'$ precipitate phase in Ni-based superalloys, Ta remains in low concentration in $\gamma$ [27]. In addition to the large atomic size of Ta, simulation suggests that there is a slightly attractive interaction between the Ta and vacancies [28]. This is also expected for the interaction between $Re$ and vacancies in the few first neighbor shells. Yet this has not been experimentally proven.

* Corresponding author.
*E-mail addresses:* f.ferraz@mpie.de (F.F. Morgado), b.gault@mpie.de (B. Gault), l.stephenson@mpie.de (L.T. Stephenson).

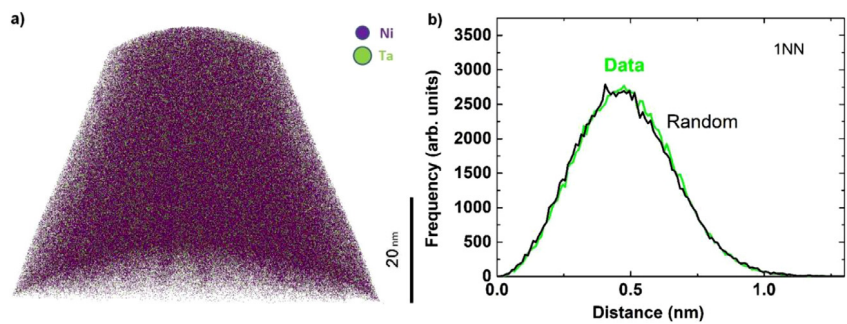

**Fig. 1.** Atom probe tomography measurement of the crept sample. a) Reconstruction of the deformed sample. b) First nearest neighbor distribution between tantalum atoms.

Here, we characterize a creep-deformed binary Ni-2 at.% Ta alloy using APT and aFIM. APT analysis reveals a random distribution of Ta within the Ni matrix. aFIM allows to image vacancies located in the first nearest-neighbor position to Ta atoms. The FIM contrast is interpreted by using density-functional theory (DFT) calculations. Our study helps to confirm the predicted interactions of vacancies with Ta.

All details regarding the preparation of the alloy can be found in ref. [26]. A solid solution single crystal alloy of Ni-2 at.% Ta was cast. A cylindrical specimen was subjected to a uniaxial compression at 1050 °C using a pneumatic compression creep testing machine at a constant applied stress of 20 MPa until a plastic strain of about 11% was reached within 2.9 h.

Needle-shaped specimens suitable for APT and FIM were prepared at the FEI Helios PFIB, which uses a xenon plasma ion source instead of the more common gallium source. The preparation procedure is discussed in [29]. Both APT and FIM measurements were performed on a LEAP 5000 XS (Cameca Instruments Inc.). The APT measurements were conducted in voltage pulsing mode using a pulse fraction of 20%, a pulse repetition rate of 200 kHz, a temperature of 40 K, and a detection rate of 0.5%. APT data processing and reconstruction were performed using the commercial software IVAS 3.8.4. The FIM measurements were done with a 99.999% purity neon as imaging gas via a manually controlled leak valve. The pressure in the analysis chamber was maintained at approximately $1 \times 10^{-6}$ mbar. The aFIM technique was performed using the APT in-situ time-of-flight mass spectrometer by superimposing a pulsed voltage on the DC voltage (pulse fraction of 20%) at the FIM working pressure. The data is extracted as an EPOS file and processed using in-house built routines within MATLAB R2019a.

We first used APT to confirm the composition of the alloy and assess the Ta-distribution. Fig. 1(a) is the tomographic reconstruction showing the Ni and Ta atoms in purple and green, respectively. The composition found was Ni 97.9 and Ta 2.1 at.%, which is very close to the nominal composition. Fig. 1(b) shows the distribution of distance to the first-nearest neighbor Ta atom within the reconstructed APT point cloud, along with a corresponding random distribution [30]. No clustering is observed.

Fig. 2(a) shows a field ion micrograph of the crept sample. The image reveals the symmetries associated with the alloy's face-centered cubic structure, and, in some of the atomic terraces, a true atomic resolution is achieved. The enlarged region corresponds to the (012) atomic planes. As we increase the voltage, the field evaporation of the surface atoms allows us to image deeper layers within the material. Fig. 2(b)-(e) shows the successive atomic planes as the topmost layers are removed nearly atom-by-atom. Some atoms are imaged larger and brighter than others. Fig. 2(e) shows a vacancy underneath one of the brighter atoms, and it can be clearly seen after a few nearby atoms are field evaporated. Even after the evaporation of the atomic plane, no atom is ever imaged where the vacancy is first observed. This confirms that it is a vacancy and not an artifact of imaging or out-of-sequence evaporation.

Assuming random site occupations, i.e. without considering any attraction between vacancies and Ta, the probability that at least one tantalum atom is directly neighboring to a vacancy would be approximately 21.5%. We observed a vacancy-tantalum complex with each measurement (10 in total). Taking each measurement as independent and assuming a random site occupation probability as a null hypothesis, the p-value of the combined observation is $2.1 \times 10^{-7}$. This small p-value suggests that the vacancy-tantalum interaction is non-random. Ni-matrix vacancies not associated with Ta (bright imaged atoms) were not observed. This observation was made on multiple occasions – see supplementary information for other examples. The micrograph in Fig. S1 was obtained using a conventional field-ion microscope. The observed vacancy could be an artifact created by the field evaporation of the brighter atom, as reported by Stiller et al. [31]. However, this artifact cannot account for all observations. For instance, Fig. S2 shows a vacancy in the plane, which is imaged first, above the brightly imaging Ta atom.

To interpret the brightness contrast between Ni and Ta, we investigated the geometric and electronic structure of the Ni (012) surface by using density functional theory (DFT) calculations in the presence of an electric field of 40 V/nm. DFT was performed in the plane-wave PAW formalism with the SPHInX code [32] using the Perdew-Burke-Ernzerhof (PBE) exchange-correlation functional with D2 van-der-Waals corrections, which allowed us to investigate the interaction with Ne adatoms. The plane-wave cutoff was 20 Ry. The Ni (012) surface was modeled in the repeated slab approach with 9 atomic layers at the theoretical lattice constant (3.465 Å). Finite electric fields of 40 V/nm were introduced with the generalized dipole correction [33]. Tantalum substitution at the surface was modeled in a 3 × 3 surface unit cell, employing a 2 × 3 × 1 offset $k$-point sampling (4 × 6 × 1 for DOS calculations). Calculations including a Ne-adlayer did not show any notable influence.

Due to the electrostatic field, the surface becomes charged by 0.3 atomic charge units (e) per surface atom. Using Hirshfeld de-composition to assign the charges to individual atoms, we find that the surface charge on this relatively open surface is located 90% on the top-layer atoms, and 10% on the second layer. Substitutional Ta relaxes outward compared to the surrounding Ni atoms by ≈ 0.1 Å due to its slightly larger atomic size, but acquires a significantly higher charge (0.6e), which is transferred from the other eight top-surface atoms (in total: 0.15e) and the three second-layer atoms directly below Ta (0.14e). The increased local electrostatic field likely affects the probability of tunneling from the Ne p-electrons into the surface, but it also enhances the local concentration of imaging gas atoms [34,35].

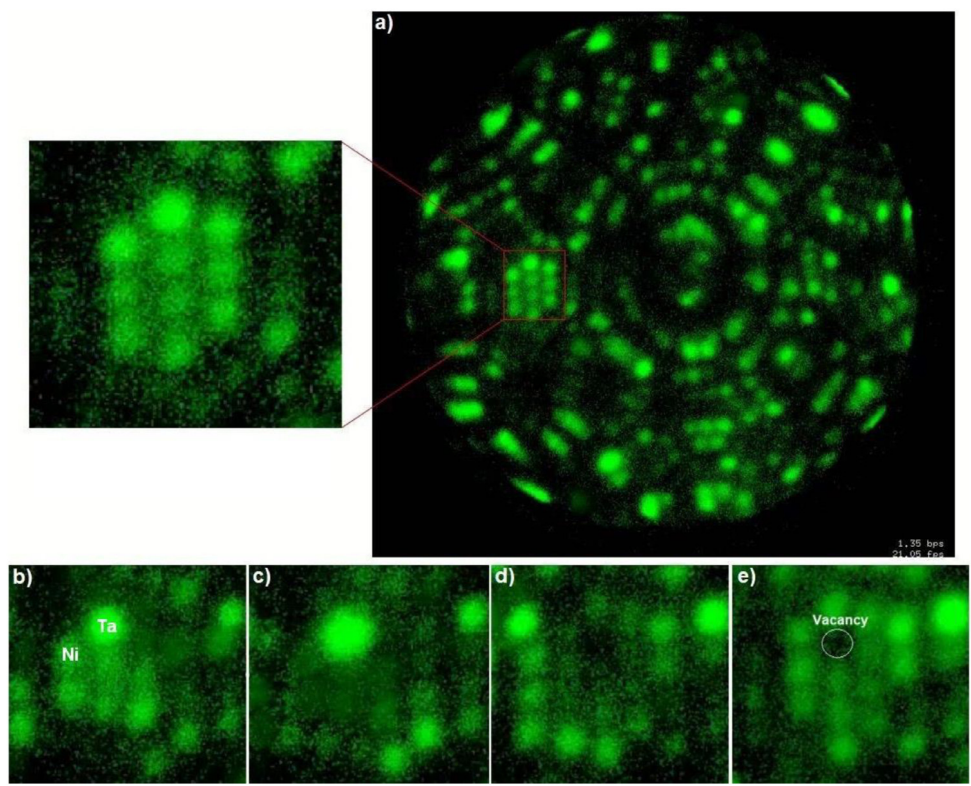

**Fig. 2.** Field ion microscopy measurement. (a) Field ion micrograph zoomed on the (012) planes. (b)-(e) shows consecutively atomic layers.

In addition to the effects associated with the electrostatic field, we also propose a significant electronic effect. As Ni contains an almost filled D-band, tunneling into Ni D-states is restricted to a small energy window up to 1 eV above the Fermi level in the spin minority channel - the spin majority D-band is completely filled even for charged surfaces. Ta, however, possesses a very high local density of unoccupied d-orbitals 0.5–3 eV above the Fermi level in both spin channels, as shown in Fig. 3. Tunneling into these orbitals will further enhance the local ionization probability and hence the FIM brightness. We refrain from a quantitative analysis at this stage because tunneling into high-lying states is damped by the larger tunneling distance required to align the Ne level with the target state. Yet, we can assume that the brightest atoms in the field-ion micrograph are Ta, while the others can be attributed to Ni (Fig. 2(b)).

Experimentally, we studied the brightness contrast with aFIM. Fig. 4(a) shows the recalculated FIM image, a two-dimensional histogram using 600,000 successive detector impact positions. The (240) planes, delineated by the red square, were selected for the chemical identification as they exhibit an imaging resolution enabling the identification of individual atoms. As aFIM is operated at a pressure of $10^{-6}$ mbar, the mass-to-charge spectrum was filtered using a similar protocol as Katnagallu et al. [13]. It uses the spatial correlations in the detection of Ne and Ni or Ne and Ta originating from the same high-voltage pulse. This protocol leads to the detection of $Ni^{2+}$ ions shortly after the disappearance of the corresponding dimmer atom in the FIM image, shown in Fig. 4(b). In comparison, the $Ta^{3+}$ ion is detected immediately after the brighter atom's field evaporation, as shown in Fig. 4(c).

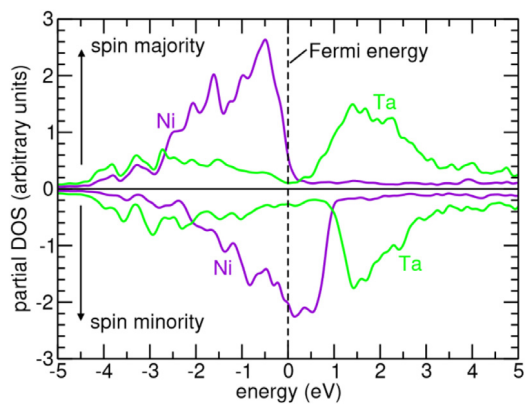

**Fig. 3.** Partial density of states for a representative Ni atom and Ta in the top surface layer of a Ni (012) surface. Positive values correspond to spin majority, negative values to spin minority channel.

In summary, we provide experimental evidence for vacancy-Ta affinity through aFIM-DFT in a crept Ni-2 at.% Ta alloy. A random distribution of Ta in Ni is first revealed by a Ta-Ta nearest neighbor distance analysis of the atom probe data. DFT calculations show that Ta atoms relax outwards compared to the surrounding Ni atoms and acquire higher electrical charges. We attribute the FIM contrast to a combination of electric and electronic factors: the charge induced local field enhances the local concentration of the gas and facilitates electron tunneling from the Ne imaging gas, which is further enhanced by the appearance of unoccupied Ta d-orbitals 1–3 eV above the Fermi level. Meanwhile, the bright contrast of Ta solute in the Ni-matrix reveals a positive correlation between Ta solutes and vacancies. Our observations support the ab-initio results of Schuwalow et al. [28], who proposed attractive solute-vacancies interactions in Ni-Ta, in the first neighbor shell. It is also in agreement with the work reported by Pleiter and Hohenemser [36], which shows some positive correlation between vacancies and large atomic radius solutes. According to Ur Rehman et al. [26], the Ta solutes provide a more potent hardening than Re for lower temperatures. The binding of vacancies by the nearly homogeneously distributed Ta atoms within the Ni-matrix could potentially slow down dislocation climb processes [37,38] involved in creep deformation, improving its creep resistance.

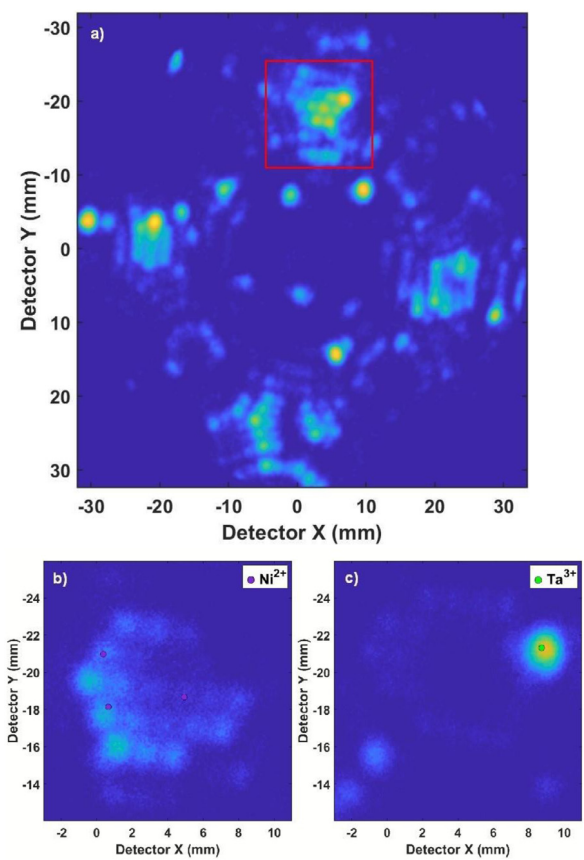

**Fig. 4.** Analytical field ion microscopy measurement. (a) The field ion micrograph with the plane (240) highlighted. (b)-(c) shows the chemical mapping of $Ni^{2+}$ and $Ta^{3+}$ ions in the enlarged region.

This work was supported by The International Max Planck Research School for Interface Controlled Materials for Energy Conversion (IMPRS-SurMat). We also appreciate the technical support of U. Tezins and A. Sturm at the APT/FIB facilities at Max-Planck-Institut für Eisenforschung. LTS & BG author acknowledges financial support from the ERC–CoG-SHINE-771,602. The authors acknowledge funding by the Deutsche Forschungsgemeinschaft (DFG) through projects A4 and B3 of the collaborative research center SFB/TR 103 "From Atoms to Turbine Blades – a Scientific Approach for Developing the Next Generation of Single Crystal Superalloys".

## Declaration of Competing Interest

The authors declare that they have no known competing financial interests or personal relationships that could have appeared to influence the work reported in this paper.

## Supplementary materials

Supplementary material associated with this article can be found, in the online version, at doi:10.1016/j.scriptamat.2021.114036.